\documentstyle[12pt]{article}
\def\jump{\vskip0.5truecm}
\def\be{\begin{equation}}
\def\ee{\end{equation}}
\def\l{\label}
\def\ba{\begin{array}}
\def\ea{\end{array}}

\def\refe#1{(\ref{#1})}

\begin{document}
\begin{titlepage}
\vspace*{-1.5cm}
\begin{center}

\hfill IC/96/67  
\\[1ex]  \hfill April, 1996

\vspace{5ex}
{\Large \bf Long range neutrino forces and 
\\[1ex] the lower bound on neutrino mass}

\vspace{3ex}
{\bf
Alexei Yu. Smirnov$^{a,b}$ and Francesco Vissani$^{a}$
}

{\it
\vspace{1ex}  ${}^a$ International Centre for Theoretical Physics, ICTP
\\[-1ex] Via Costiera 11, I-34013 Trieste, Italy
}

{\it
\vspace{1ex}  ${}^b$ Institute for Nuclear Research,
\\[-1ex] Russian Academy of Sciences,
\\[-1ex] 117312 Moscow,  Russia
}

\vspace{6ex}
{ABSTRACT}
\end{center}

\begin{quotation}

Stellar objects (including our Sun,  other stars of main 
sequence, white dwarfs, neutron stars  etc.) 
contain strongly degenerate low energy    
sea of neutrinos (in neutron stars) or antineutrinos. 
The presence of this sea leads due to Pauli principle 
and thermal effects to  
effective blocking of the long-range neutrino 
forces. 
This blocking can resolve the problem of the unphysically large value of 
the self-energy of stars stipulated by many body  long-range  
neutrino interactions. As a consequence no lower bound 
on the neutrino mass is inferred, in contrast with the  
statement by Fischbach.  

\end{quotation}
\end{titlepage}
\vfill\eject

\section{The self-energy paradox} 

The exchange of massless neutrinos leads to  
long range forces. The  potential of interaction of two 
particles $f$ due to these forces equals \cite{Feinberg} 
\be
V^{(2)}(r) = \frac{G_F^2 a_f^2}{4 \pi^3 r^5} = \frac{2}{\pi r} 
\left(\frac{G_F a_f}{2\sqrt{2}\pi r^2}\right)^2, 
\l{poten2}
\ee
where $G_F$ is the Fermi coupling constant, $a_f$ is  
the weak vector charge of particle $f$, and 
$r$ is the distance between the particles. 
For neutrons, electrons and protons one has  
$a_n = -1/2$,  $a_e = 1/2 + 2\sin^2 \theta$,  
$a_p = 1/2 + 2\sin^2 \theta$.  
Neutrino exchange leads also to many body interactions 
which can be represented in  the $k$-body case   
as the  neutrino loop interacting with 
$k$ currents of $f$ (ring diagrams). 
Inclusion of one additional particle $f$ adds the factor 
\be
\frac{G_F a_f}{2\sqrt{2}\pi r^2}  
\l{factor}
\ee 
to the potential, and for $r \sim R$, the typical size of   
star, this factor is extremely small.  

It is claimed \cite{Fischbach}, that  
in spite of the smallness of factor (\ref{factor})  
the many-body neutrino forces 
become dominating in 
stars, where the number of particles, $N,$ can exceed $10^{57}$.  
The contribution of the $k$-body interactions 
to the self-energy of star, $W^{(k)},$ can be represented as 
\be 
W^{(k)} = 
U^{(k)} \cdot C^N_k \approx 
\frac{1}{k!} \cdot U^{(k)}  N^k,
\l{self}
\ee
where $U^{(k)}$ is the $k$-body potential averaged over the volume 
of  star and  
$C^N_k = \frac{N!}{k!(N - k)!}$ is the number of combinations 
of $k$ particles among $N$ particles. It is this combinatoric 
factor, first discussed in \cite{Primakoff},  which leads to  the 
dominance of the 
many-body 
interactions. The second equality is valid for values of $k \ll N$,  
when $C^N_k \approx N^k/k!$. 
The ratio of contributions to the self energy 
from $k + 2$ and $k$ body interactions equals  
\be
\left|\frac{W^{(k + 2)}}{W^{(k)}}\right| \approx \xi^2  ,
\l{ratio}
\ee
where according to (\ref{factor}) and (\ref{self}) \cite{Fischbach}
the series parameter $\xi$ is 
\be
\xi \equiv  
\frac{a_f}{2\sqrt{2}\pi e} \frac{G_F N}{R^2} = 
\frac{\sqrt{2} a_f}{3 e}~ 
G_F n_f~ R 
\label{xi}
\ee
($e = 2.718...$). Here $R$ is the radius of 
star and $n_f$ is the average density of particles. 
Notice that $\xi$ is determined essentially by the width of matter 
in star. For neutron star with 
$R \approx 10$ km and $N \approx 10^{57}$ one finds 
the series parameter $\xi \approx 10^{13}$. 
Thus the contribution to self-energy enormously 
increases with multiplicity of interaction. For $k = 8$ 
one finds $W^{(8)} \approx 3 \cdot 10^{11} M_{NS}$,  i.e. formally 
the  self-energy exceeds the mass of the  
star \cite{Fischbach}. 
There are some other aspects of the paradox: the   
self energy changes the sign if two more neutrons are added 
to star, or when microscopically 
small changes of radius of  star occur.

It is claimed in  \cite{Fischbach} that the only way to resolve the paradox 
is to suggest that  neutrino has non-zero mass $m_{\nu}$. 
In this case the effective radius of interaction becomes 
$r \sim 1/m_{\nu}$ and in the series parameter 
one should substitute $R$ by $1/m_{\nu}$:  
\be
\xi_m = 
\frac{\sqrt{2} a_f}{3 e}~ 
\frac{G_F n_f}{m_{\nu}} .   
\label{xim}
\ee
This means that the neutrons within the radius $1/m_{\nu}$ 
only contribute to the many-body potential. From the condition
\be
\xi_m < O(1) , 
\ee 
(more precisely: the exchange energy 
inside a given volume should be smaller 
than the mass of that volume)  
one finds the lower bound on neutrino mass 
$m_{\nu} > 0.4$ eV \cite{Fischbach}. If correct,  
this result has a number of 
very important consequences.\jump
 
In this paper we describe another mechanism of  
suppression of the long-range interactions. It  
can systematically lead to  $\xi < 1$, so that 
many body forces do not dominate in the self-energy, and no lower 
bound on the 
neutrino mass is inferred. The suppression effect originates from the 
neutrino sea which exists in  stars, 
in contrast with a statement done in reference \cite{Fischbach}.\jump

\section{Neutrino sea} 

The incorrect statement about neutrino 
sea in \cite{Fischbach} follows 
from the observation that the neutrino 
cross-section at low energies 
are extremely small, so that the neutrinos freely escape from 
the star. However, as it has been shown by Loeb \cite{Loeb},  
the low energy neutrinos are trapped in the stars due to refraction. 
Refraction effect is stipulated by coherent neutrino interaction 
with particles of medium. 
For low energies the refraction index, $G_F n_f/E$, becomes 
of the order 1 and 
the effect of the complete inner reflection takes place.  
Neutrinos have circular orbits inside the star \cite{Loeb}. 
Equivalently, one can describe the effect in terms of  
potential created by particles of medium.

For definiteness let us consider the neutron star. 
For neutrinos, the  star can be 
viewed as the potential well with the depth: 
\be
V \approx \frac{1}{2}G_F n_n = 10~ {\rm eV} 
\left(\frac{n_n}{10^{37} {\rm cm}^{-3}}\right) .
\ee
The potential is attractive for neutrinos which have the 
positive weak charge and repulsive for antineutrinos. Therefore  
produced  antineutrinos leave the star, whereas the 
neutrinos turn out to be trapped. Neutrinos fill in the levels of 
the potential well of the star. As the result the  sea is strongly 
degenerate \cite{Loeb}. The energy distribution of neutrinos 
can be approximated by the thermal Fermi-Dirac distribution 
\be
n_\nu (E) = \frac{1}{1 + \exp{\frac{E - \mu}{T}}} 
\l{dist}
\ee
with chemical potential  
$$
\mu \approx V = a_f G_F n_f
$$  
and the effective temperature $T \ll \mu$. 
(The effective temperature describes  the degree of filling 
in the levels of the well.) For neutron star we get numerically 
$\mu  \approx 50$ eV.

The total density of the neutrinos \cite{Loeb} equals: 
\be
\int \frac{d^3 p}{(2\pi)^3} n(p) \approx \frac{V^3}{6\pi^2} \approx 
3 \cdot 10^{17} {\rm cm}^{-3} . 
\l{dens}
\ee
Notice, this number is much  smaller than that of neutrons: 
$n_{\nu}/n_N \sim 10^{-20},$ and the energy stored in these neutrinos  
is too  small to influence  the dynamics of star.\jump

The degenerate sea exists  starting from early 
stages of  evolution in all stars. In the Sun the potential $V$ is 
created by 
the interactions of neutrinos with protons, neutrons and electrons 
$V  = \sum a_f G_F n_f \sim 10^{-13}$ eV. It has an opposite sign to 
that in the neutron star so that the Sun has the antineutrino sea. 
(Also white dwarfs have antineutrino sea, and only in the process 
of neutronization in the  
protoneutron star, when $n_n$ becomes bigger than $2 n_p$, 
the potential changes the sign).   
According to (\ref{dens}) the concentration of neutrinos 
in the sea of the Sun equals $n_{\nu} \sim 10^{-28}$ cm$^{-3}$ 
and the total number of these neutrinos is $10^5$. 
However even this small number can be sufficient 
to resolve the energy paradox. 

There is a number of processes which can 
contribute to creation of the neutrino sea. 
For example, in the neutron stars 
the neutronisation itself can lead to strongly degenerate spectrum of 
neutrinos. One can consider also the neutrino pair 
bremsstrahlung in the collisions of neutrons, URCA processes, 
neutrino production by the weak field etc.. 
However the most efficient mechanism of production 
of very low energy neutrinos is the one related to the long-range 
many-body forces themselves \cite{in-prep}. 
Indeed, as we described before,  many body forces are 
induced by the neutrino loops which couple  
with k-currents of particles of medium. 
If in such a diagram 
one of the propagators of neutrino between points 
$x$ and $y$ is substituted by emission of neutrino in 
$x$ and absorption of neutrino in $y$ 
(and vice versa) one gets the 
diagram for neutrino pair production. 
Obviously, this process has  
the same combinatoric enchancement 
as the self-energy contributions: 
$\sigma^{(k)} \propto  k U^{(k)} C^{(k)}_{N}$. 
So that the efficiency of the neutrino pair production 
increases together with the self-energy.\jump

\section{Blocking effect} 

Let us show now that  presence of the 
degenerate neutrino sea can lead to the 
blocking of the 
long range forces stipulated by the neutrino exchange. 
For this let us consider the propagator of massless neutrino 
in the  degenerate neutrino sea: 
\be
\begin{array}{l}
S(x - y)=\displaystyle\int\frac{d k^4}{(2\pi)^4}\ 
e^{-ik(x-y)}\  \hat k \times\\
\ \ \displaystyle \left[ \frac{i}{k^2 + i\epsilon} 
-2\pi  \delta (k^2)\ \theta ( k_0)\ n_{\nu}(k_0)
-2\pi  \delta (k^2)\ \theta (-k_0)\ n_{\bar\nu}(-k_0) \right] \ .
\end{array}
\l{prop1}
\ee
The thermal distribution $n_{\nu}(E)$ is given in  (\ref{dist});
for $n_{\bar\nu}(E)$ one should substitute
$\mu\to -\mu$ in $n_{\nu}(E).$  
The first term is the vacuum propagator. 
The second term corresponds to real neutrinos of the sea. 
This term describes the process, in which  instead of the exchange 
of a virtual particle between two points $x$, $y$, one has the 
absorption of neutrino in the point $x$ and the emission of  neutrino 
in the point $y$ and vice-versa; 
similar consideration is applied to the last term in (\ref{prop1}).

Let us find  the potential due to neutrino exchange
using Schwinger formalism \cite{Schwinger}.
According to this formalism 
the neutrons can be treated  as static classical sources 
of quantized neutrino field. Then  the   
multi-body potential due to neutrino loops,  
$V(\bar x_1,\bar x_2,\bar x_3,...)$, can be found by calculation   
of the integral over the energy 
of 
$$
tr[\gamma_0 P_L\ S(E,\bar x_1-\bar x_2)\ 
 \gamma_0 P_L\ S(E,\bar x_2-\bar x_3)\ ...] .
$$
Here $S(E,\bar x)$ is the neutrino propagator being a
function of the energy and of the position, $P_L$ is the
chirality projector, and $tr$ is the trace over the spinorial indices.
Using equation \refe{prop1} we find the propagator:
\be
\begin{array}{rcl}
S(E,\bar x)      &=& (-E \gamma_0 + i\partial_i \gamma_i) 
\Delta(E,\bar x) ,  \\[1ex]
\Delta(E,\bar x) &=& \displaystyle
-\frac{i}{4\pi r} 
\left[\
(1-n_\nu (E))\ \exp(i E r) +
n_\nu (E)\ \exp(-i E r) \ \right] \ . 
\end{array}
\l{prop3}
\ee
Substituting the propagators of this type in Schwinger formula  
\cite{Schwinger} 
we obtain the  
the two-body potential:
\be
V^{(2)}_{\mu}(r)=\frac{(G_F a_f)^2}{4 \pi^3 r^5} 
\left[
\cos(2\mu r) + \mu r \ \sin(2\mu r) + 
\frac{\cos(2\mu r)}{2}  {\cal F}(2 \pi T r)
\right]\ \frac{2 \pi T r }{\sinh 2\pi  T r}
\l{twobod}
\ee
instead of \refe{poten2}. In 
(\ref{twobod}) 
${\cal F}(x)=x \coth(x) -1$; 
this function behaves like  
$x$  
for large $x,$ whereas for small $x$ one has  
${\cal F}(x)=x^2/3 + {\cal O}(x^4).$ 
For zero temperatures the potential 
(\ref{twobod}) reduces to 
\be
V^{(2)}_{\mu}(r)=\frac{(G_F a_f)^2}{4 \pi^3 r^5} 
\left[
\cos(2\mu r) + \mu r \ \sin(2\mu r)\right]\ . 
\l{twobody}
\ee
This result coincides with the one which can be obtained from the 
calculations by  Horowitz and Pantaleone
\cite{Pantaleone}.
The potential \refe{twobod} 
shows the blocking effect: the presence 
of the neutrino sea results in  fastly oscillating factors
$\cos(2 \mu r)$ and $\sin(2 \mu r)$ as well as in 
an exponential damping 
of the long range  neutrino forces. 

The results \refe{prop3}, \refe{twobod}, \refe{twobody} 
correspond to thermal equilibrium. 
Actually   the neutrino sea 
distribution  is non-thermal already due to the fact that the 
antineutrinos freely escape from the star. 
In this connection let us consider extreme situation 
of the complete antineutrino stripping.   
It corresponds to the omission
of the last term in propagator \refe{prop1}. 
In this case  the potential acquires an additional contribution:  
$V^{(2)}_{stripped} =  V^{(2)}_{\mu} + 
\Delta V^{(2)}$, where   
\be
\Delta V^{(2)}(r)=-\frac{(G_F a_f)^2}{8 \pi^3 r^5} 
\sum_{n=1}^\infty [-\exp(-\mu/T)]^n\ 
\frac{[n/(2 T r)]^2}{(1+[n/(2 T r)]^2)^2} \ .  
\l{dtwobod}
\ee 
For strongly degenerate sea  ($T<<\mu$)   
it is enough to  use 
the first term in the series (\ref{dtwobod})),  
so that 
$$
\Delta V^{(2)}(r) \propto \frac{(G_F a_f)^2}{8 \pi^3 r^5} 
\exp(-\mu/T) . 
%\frac{[1/(2 T r)]^2}{(1+[1/(2 T r)]^2)^2} \ ,  
%\l{dtwobod}
$$ 
The contribution $\Delta V^{(2)}$   
has no oscillatory behaviour; it reflects  
the incompleteness of filling in 
the low energy levels $E \rightarrow 0$ (in accordance with  
(\ref{dist})). 
Strong blocking effects  
requires strong degeneracy of the sea: 
$T<<\mu$. For example, the  many 
body interactions in the neutron star are suppressed (so that 
$\xi < 1$), if  
$\exp(-\mu/T) < 10^{-13}$,  or 
$\mu/T > 30$.  

Thus,  the effect of the term (\ref{dtwobod}) is suppressed,  if 
the sea is strongly enough degenerate. The 
contribution of the term 
(\ref{twobod}) to the self-energy
is suppressed due to fast oscillations. 
Indeed, the integration over the volume of  star,  
the motion of particles as well as macroscopic 
motion of medium, fluctuations 
of density and radius of the star etc.,  lead 
to strong cancellation of contributions from distances 
$r > 1/ \mu$. Effectively this gives the cut of the long-range 
forces at $r \sim 1/\mu$. 

Similar oscillating factors appear in  many body 
potentials \cite{in-prep}. 

Thus, if in the series parameter $\xi$ of Eq.\ (\ref{xi})  
the radius $R$ is substituted by $1/\mu,$  
one gets for the parameter in presence of 
the degenerate sea: 
\be 
\xi_{\mu} =  
\frac{\sqrt{2}}{3 e}~ 
\frac{a_f G_F n_f}{\mu} =  \frac{\sqrt{2}}{3 e} < 1 . 
\label{xi1}
\ee
The series parameter becomes smaller 
than one. Moreover, it does not depend on properties of the star: 
density, radius, chemical composition. 
This shows the {\it universality} of the 
mechanism; it works for any stars. 
Due to $\xi_{\mu} < 1$ and presence of  additional factor 
$1/k!$ in $W^{(k)}$ the many body forces  never exceed the 
lowest contributions, and this  solves the  paradox.

In conclusion, let us us outline the dynamics of blocking.  
Let us start by very diluted  stellar object 
--- star at early stages of evolution, 
for which even without blocking one has $\xi < 1$.  
With increase of density  the self-energy increases. 
However, simultaneously the 
potential of medium increases, and  
in the same degree the efficiency of low energy neutrino 
pair production increases. The neutrinos (or antineutrinos ) 
freely escape from the star with velocity of light and antineutrinos 
(or neutrinos) fill in the potential well. The degenerate 
neutrino sea leads to Pauli 
blocking and thermal damping
of the small energy neutrino exchanges and therefore 
to blocking of the long-range forces.  Moreover due to 
Pauli principle the sea will block further production 
of neutrino pairs. 
%New neutrinos can be produced if their 
%energy exceeds $1/\mu$ which means that the production 
%becomes a short-range process with negligibly small cross-section.  

\newpage

\noindent
{\Large\bf Acknowledgements.}
\jump

The authors  would like to thank  
E. Kh. Akhmedov and S. Zane
for useful discussions.\jump

\end{document}